\documentclass[prd,showpacs,amsmath,amssymb,twocolumn]{revtex4}
\usepackage{graphicx,color}
\begin{document}

\title{{$X(1576)$ and the Final State Interaction Effect}}

\author{Xiang Liu}\email{xiangliu@pku.edu.cn}
\author{Bo Zhang}\author{Lei-Lei Shen}
\author{Shi-Lin Zhu}
\affiliation{Department of physics, Peking University,
Beijing, 100871, China}

\vspace*{1.0cm}

\date{\today}
\begin{abstract}
We study whether the broad peak $X(1576)$ observed by BES
Collaboration arises from the final state interaction effect of
$\rho(1450,1700)$ decays. The interference effect could produce an
enhancement around $1540$ MeV in the $K^+K^-$ spectrum with typical
interference phases. However, the branching ratio $B[J/\psi\to
\pi^{0}\rho(1450,1700)]\cdot B[\rho(1450,1700)\to K^{+}K^{-}]$ from
the final state interaction effect is far less than the experimental
data.
\end{abstract}

 \maketitle

Recently BES Collaboration reported a broad signal $X(1576)$ in
the $K^{+}K^{-}$ invariant mass spectrum in the $J/\psi\to
\pi^{0}K^{+}K^{-}$ channel \cite{1576-BES}. Its quantum number and
mass are $J^{PC}=1^{--}$ and
$m=(1576^{+49}_{-55}(\mathrm{stat})^{+98}_{-91}(\mathrm{syst}))
-i(409^{+11}_{-12}(\mathrm{stat})^{+32}_{-67}(\mathrm{syst}))$ MeV
respectively . The branching ratio is $B[J/\psi\to
X(1576)\pi^{0}]\cdot B[X(1576)\to K^{+}K^{-}]=(8.5\pm
0.6^{+2.7}_{-3.6})\times 10^{-4}$. If one ignores the tiny isospin
violation effect and assume both isospin and G-parity are good
quantum numbers, $X(1576)$ is an isovector with even G-parity.
However, the most notable character of $X(1576)$ is its extremely
large width around 800 MeV, which motivated theoretical
speculations that it could be a $K^{*}(892)-\kappa$ molecular
state \cite{Guo-1576}, tetraquark \cite{Lipkin-1576,1576-QSR},
diquark-antidiquark bound state \cite{Ding-1576,zhang}.

The lowest scalar meson $\sigma$ is also very broad \cite{PDG}.
After decades of experimental and theoretical investigations, the
underlying structure of the $\sigma$ meson is still elusive now.
Although exotic theoretical interpretations such as treating
X(1576) as a tetraquark is quite interesting, one must answer
where are those partner states of X(1576) in the same tetraquark
multiplet. On the other hand, it will be worthwhile and equally
interesting to explore whether such a broad signal could be
produced by more conventional schemes like the final state
interaction (FSI) effect.

It's interesting to note that there are two broad overlapping
resonances $\rho(1450)$ and $\rho(1700)$ with the same quantum
number as X(1576) around 1600 MeV. Their widths are about 147 MeV
and 250 MeV respectively \cite{PDG}. The FSI effect sometimes plays
a very important role in some processes \cite{FSI}. Hence, we want
to take a look at the FSI effect in the $\rho(1450,1700)\to
K^{+}K^{-}$ decays and explore whether the FSI effect may produce a
similar broad signal in this work.

\begin{figure}[htb]
\begin{center}
\begin{tabular}{cc}
\scalebox{0.7}{\includegraphics{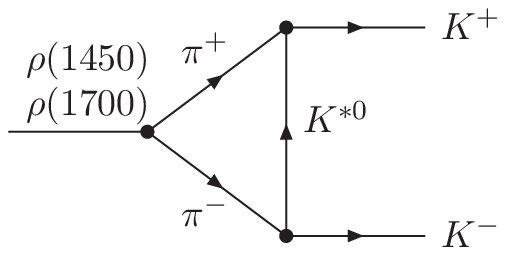}}&
\scalebox{0.7}{\includegraphics{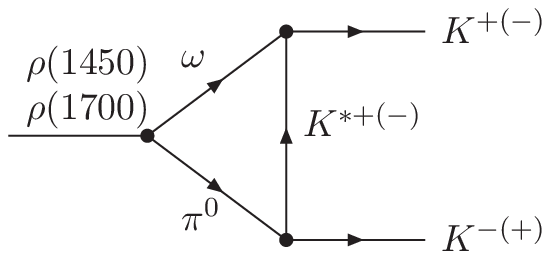}}\\
(a)&(b)
\\
\scalebox{0.7}{\includegraphics{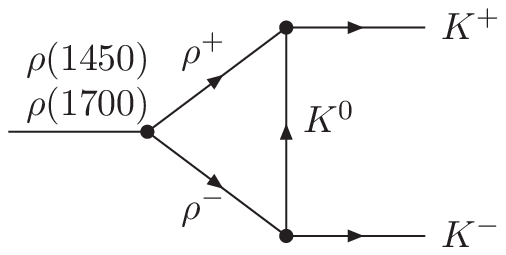}}&\scalebox{0.7}{\includegraphics{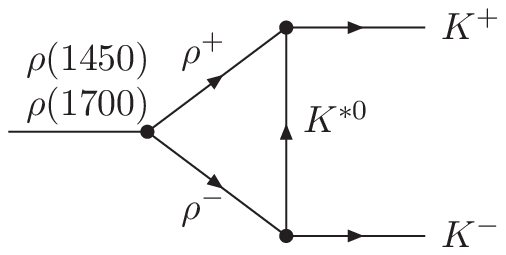}}\\
(c)&(d)\\
\scalebox{0.7}{\includegraphics{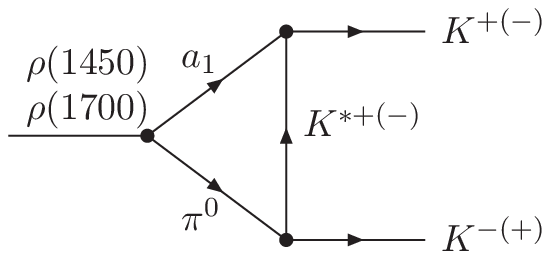}}&\\
(e)&\\
\end{tabular}
\end{center}
\caption{Some possible intermediate states contributing to the
$\rho(1450,1700)\to K^{+}K^{-}$ decays.}\label{KK}
\end{figure}

The intermediate states $\pi^{+}\pi^{-}$, $ \omega\pi^{0}$,
$\rho^{0}\eta$, $\rho^{+}\rho^{-}$, $a_{1}(1260)\pi^0$ contribute to
$\rho(1450,1700)\to K^{+}K^{-}$ as shown in Fig. \ref{KK}. The
$K^{*+}K^{-}+c.c.$ intermediate state also contributes to
$\rho(1700)\to K^{+}K^{-}$. The above intermediate states are of
four types: $P+P$, $P+V$, $V+V$ and $A+P$, where $P$, $V$ and $S$
denote pseudoscalar, vector and axial vector mesons respectively.

The effective Lagrangians related to our following calculation
read
\begin{eqnarray}
\mathcal{L}_{V_{1}\to P_{1}P_{2}}&=&ig_{1}(P_{1}{\stackrel{\leftrightarrow}{\partial}}P_{2})V^{\nu},\\
\mathcal{L}_{V_{1}\to
V_{2}P_{1}}&=&g_{2}\epsilon_{\mu\nu\alpha\beta}V_{1}^{\mu}\partial^{\nu}P_{1}\partial^{\beta}V_{2}^{\alpha},
\\
\mathcal{L}_{V_{1}\to
V_{2}V_{3}}&=&ig_{3}\Big\{V_{1}^{\mu}(\partial_{\mu}V_{2}^{\nu}V_{3\nu}-V_{2}^{\nu}\partial_{\mu}V_{3\nu})\nonumber\\&&
+(\partial_{\mu}V_{1\nu}V_{2}^{\nu}-V_{1\nu}\partial_{\mu}V_{2})V_{3}^{\mu}\nonumber\\&&+
V_{2}^{\mu}(V_{1}^{\nu}\partial_{\mu}V_{3\nu}-\partial_{\mu}V_{1\nu}V_{3}^{\nu})
\Big\},\\
\mathcal{}L_{S\to P_{1}V_{1}}&=&g_{4}P_{1}V_{1}^{\mu}S_{\mu},
\end{eqnarray}
where $g_{i}$'s denote the coupling constants. $P_{1,2}$,
$V_{1,2,3}$ and $S$ respectively represent pseudoscalar, vector
and axial vector fields.

Using the Cutkosky rule, one obtains the the absorptive
contribution to the process of $\rho(1450,1700)\to
\pi^{+}(p_1)\pi^{-}(p_{2})\to \rho^{+}(p_{3})\rho^{-}(p_{4})$ in
Fig. \ref{KK} (a) {\small{\begin{eqnarray}
&&\mathbf{Abs}^{(a)}[\pi^{+}\pi^{-},K^{*0}]\nonumber\\
&&=\frac{1}{2}\int\frac{d^{3}p_{1}}{(2\pi)^{3}2E_{1}}
\frac{d^{3}p_{2}}{(2\pi)^{3}2E_{2}}(2\pi)^{4}\delta^{4}(M-p_{1}-p_{2})
\nonumber\\&& \times
[ig_{_{\rho(1450)\pi\pi}}(p_{1}-p_{2})\cdot\varepsilon] [-ig_{_{\pi
K^{*}K}}(p_{1\mu}+p_{3\mu})]\nonumber\\&&\times[-ig_{_{\pi
K^{*}K}}(p_{2\nu}+p_{4\nu})]
\bigg[-g^{\mu\nu}+\frac{q^{\mu}q^{\nu}}{m_{K^{*}}^{2}}\bigg]\nonumber\\&&
\times\bigg[\frac{i}{q^2
-m_{K^{*}}^{2}}\bigg]\mathcal{F}^{2}(m_{K^{*}},q^2).\label{eq1}
\end{eqnarray}}}

The amplitude corresponding to the process of $\rho(1450,1700)\to
\omega(p_{1})\pi^{0}(p_{2})\to K^{+}(p_{3})K^{-}(p_{4})$ can be
written as {\begin{eqnarray}
&&\mathbf{Abs}^{(b)}[\omega\pi^{0},K^{*+}]\nonumber\\
&&=\frac{1}{2}\int\frac{d^{3}p_{1}}{(2\pi)^{3}2E_{1}}
\frac{d^{3}p_{2}}{(2\pi)^{3}2E_{2}}(2\pi)^{4}\delta^{4}
(M-p_{1}-p_{2}) \nonumber\\&&
\times[ig_{_{\rho(1450)\omega\pi}}\epsilon_{\kappa\lambda\beta\nu}p_{1}^{\kappa}\varepsilon^{\beta}p_{2}^{\nu}]
[ig_{_{KK^{*}\omega}}\epsilon_{\alpha\xi\sigma\gamma}p_{1}^{\alpha}
q^{\sigma}]\nonumber\\
&&\times i[g_{_{KK^{*}\pi}}(p_{2\mu}+p_{4\mu})]
\bigg[-g^{\mu\gamma}+\frac{q^{\mu}q^{\gamma}}{m_{K^{*}}^{2}}\bigg]
\nonumber\\&& \times
\bigg[-g^{\lambda\xi}+\frac{p_{1}^{\lambda}p_{1}^{\xi}}{m_{\omega}^{2}}\bigg]
\bigg[\frac{i}{q^2
-m_{K^{*}}^{2}}\bigg]\mathcal{F}^{2}(m_{K^{*}},q^2).
\end{eqnarray}}
The amplitudes of $\rho(1450,1700)\to\rho^{0}\eta\to K^{+}K^{-}$
and $\rho(1700)\to K\bar{K}^{*}+c.c.\to K^{+}K^{-}$ can be
obtained by replacing the coupling constants and masses in the
above formula. $\rho$, $\omega$ and $\phi$ are the exchanged
mesons between $K$ and $\bar{K}^{*}$.

For the processes of
$\rho(1450,1700)\to\rho^{+}(p_{1})\rho^{-}(p_{2})\to
K^{+}(p_{3})K^{-}(p_{4})$, the exchanged mesons are $K^{0}$ and
$K^{*0}$. Thus the relevant amplitudes are
\begin{eqnarray}
&&\mathbf{Abs}^{(c)}[\rho^{+}\rho^{-},K^{0}]\nonumber\\
&&=\frac{1}{2}\int\frac{d^{3}p_{1}}{(2\pi)^{3}2E_{1}}
\frac{d^{3}p_{2}}{(2\pi)^{3}2E_{2}}(2\pi)^{4}\delta^{4}
(M-p_{1}-p_{2})\nonumber\\&&\times\Big\{ig_{\rho(1450)\rho\rho}[\epsilon\cdot(p_{1}-p_{2})g_{\mu\nu}-\epsilon_{\mu}
(2p_{1}+p_{2})_{\nu}\nonumber\\&&+\epsilon_{\nu}(2p_{2}+p_{1})_{\mu}]\Big\}[ig_{_{\rho
KK}}(q_{\alpha}+p_{3\alpha})]\nonumber\\&&\times[ig_{_{\rho
KK}}(q_{\beta}-p_{4\beta})]\bigg[-g^{\nu\alpha}+\frac{p_{1}^{\nu}p_{1}^{\alpha}}{m_{\rho}^{2}}\bigg]
\nonumber\\&& \times
\bigg[-g^{\mu\beta}+\frac{p_{2}^{\mu}p_{2}^{\beta}}{m_{\rho}^{2}}\bigg]
\bigg[\frac{i}{q^2 -m_{K}^{2}}\bigg]\mathcal{F}^{2}(m_{K},q^2),
\end{eqnarray}
and {\small{\begin{eqnarray}
&&\mathbf{Abs}^{(d)}[\rho^{+}\rho^{-},K^{*0}]\nonumber\\
&&=\frac{1}{2}\int\frac{d^{3}p_{1}}{(2\pi)^{3}2E_{1}}
\frac{d^{3}p_{2}}{(2\pi)^{3}2E_{2}}(2\pi)^{4}\delta^{4}
(M-p_{1}-p_{2})\nonumber\\&&\times\Big\{ig_{\rho(1450)\rho\rho}
[\epsilon\cdot(p_{1}-p_{2})g_{\mu\nu}-\epsilon_{\mu}
(2p_{1}+p_{2})_{\nu}\nonumber\\&&+\epsilon_{\nu}(2p_{2}+p_{1})_{\mu}]\Big\}
[ig_{\rho
KK^{*}}\epsilon_{\alpha\beta\kappa\gamma}p_{1}^{\alpha}q^{\kappa}]\nonumber\\&&\times\Big[
ig_{\rho
KK^{*}}\epsilon_{\xi\lambda\delta\zeta}p_{2}^{\xi}q^{\delta}\Big]
\bigg[-g^{\nu\beta}+\frac{p_{1}^{\nu}p_{1}^{\beta}}{m_{\rho}^{2}}\bigg]
\nonumber\\&& \times
\bigg[-g^{\mu\lambda}+\frac{p_{2}^{\mu}p_{2}^{\lambda}}{m_{\rho}^{2}}\bigg]
\bigg[-g^{\gamma\zeta}+\frac{q^{\gamma}q^{\zeta}}{m_{K^{*}}^{2}}\bigg]
\nonumber\\&& \times\bigg[\frac{i}{q^2
-m_{K^{*}}^{2}}\bigg]\mathcal{F}^{2}(m_{K^{*}},q^2).
\end{eqnarray}}}

The decay amplitude for the process $\rho(1450,1700)\to
a_{1}(1260)(p_{1})\pi^{0}(p_{2})\to K^{+}(p_{3})K^{-}(p_{4})$ is
\begin{eqnarray}
&&\mathbf{Abs}^{(e)}[a_{1}\pi^{0},K^{*+}]\nonumber\\
&&=\frac{1}{2}\int\frac{d^{3}p_{1}}{(2\pi)^{3}2E_{1}}
\frac{d^{3}p_{2}}{(2\pi)^{3}2E_{2}}(2\pi)^{4}\delta^{4}
(M-p_{1}-p_{2})\nonumber\\&&\times[ig_{_{\rho
a_{1}\pi}}\epsilon^{\mu}][ig_{_{a_{1}K^{*}K}}][-ig_{_{\pi
K^{*}K}}(p_{2\beta}+p_{4\beta})]\nonumber\\&&\times
\bigg[-g_{\mu\alpha}+\frac{p_{1\mu}p_{1\alpha}}{m_{a_{1}}^{2}}\bigg]
\bigg[-g^{\beta\alpha}+\frac{q^{\beta}q^{\alpha}}{m_{K^{*}}^{2}}\bigg]
\nonumber\\&& \times\bigg[\frac{i}{q^2
-m_{K^{*}}^{2}}\bigg]\mathcal{F}^{2}(m_{K^{*}},q^2).\label{eq2}
\end{eqnarray}

In the above eqs. (\ref{eq1})-(\ref{eq2}),
$\mathcal{F}(m_{i},q^2)$ etc denotes the form factors which
compensate the off-shell effects of mesons at the vertices and are
written as \cite{HY-Chen,FF}
\begin{eqnarray}
\mathcal{F}(m_{i},q^2)=\bigg(\frac{\Lambda^{2}-m_{i}^2
}{\Lambda^{2}-q^{2}}\bigg)^{n},
\end{eqnarray}
where $\Lambda$ is a phenomenological parameter. As $q^2\to 0$ the
form factor becomes a number. If $\Lambda\gg m_{i}$, it becomes
unity. As $q^2\rightarrow\infty$, the form factor approaches to
zero. As the distance becomes very small, the inner structure
would manifest itself and the whole picture of hadron interaction
is no longer valid. Hence the form factor vanishes and plays a
role to cut off the end effect. The expression of $\Lambda$ is
\cite{HY-Chen}
\begin{eqnarray}
\Lambda(m_{i})=m_{i}+\alpha \Lambda_{QCD},\label{parameter}
\end{eqnarray}
where $m_{i}$ denotes the mass of exchanged meson and $\alpha$ is
a phenomenological parameter. Although we use $\Lambda_{QCD}=220$
MeV, the range of $\Lambda_{QCD}$ can be taken into account
through the variation of the parameter $\alpha$. In this work, we
adopt the monopole form factor
$\mathcal{F}(m_{i},q^2)={(\Lambda^{2}-m_{i}^2)
}/{(\Lambda^{2}-q^{2})}$, where $\alpha$ is of order unity and its
range is around $0.8<\alpha<2.2$ \cite{HY-Chen}.

Let's first focus on the ratio
\begin{eqnarray}
&&{R}^{AB,C}\nonumber\\&&=\frac{\big|\mathbf{q}_{KK}\big|\big|\mathcal{M}[\rho(1450,1700)\to
A+B\to
K^{+}K^{-}]\big|^{2}}{\big|\mathbf{q}_{AB}\big|\big|\mathcal{M}[\rho(1450,1700)\to
A+B]\big|^{2}},\nonumber\\
\end{eqnarray}
where $A$ and $B$ are possible intermediate states. $C$ denotes
the exchanged meson for the scattering process of $A$ and $B$
mesons. $\mathbf{q}_{KK}$ and $\mathbf{q}_{AB}$ are the decay
momenta corresponding to $\rho(1450,1700)\to A+B\to K^{+}K^{-}$
and $\rho(1450,1700)\to A+B$ respectively. The uncertainty from
the vertex of $\rho(1450,1700)\to A+B$ can be eliminated
\cite{zhao}. ${R}^{AB,C}$ depends on the masses of
$\rho(1450,1700)$. Because $\rho(1450,1700)$ are broad, this ratio
gives us information on the evolution of $R^{AB,C}$ with the
masses of $\rho(1450,1700)$ mesons.

Using $\Gamma(K^{*0}\to K\pi)=50.3$ MeV and $\Gamma(\phi\to
K^{+}K^{-})=2.09$ MeV \cite{PDG}, we obtain
$g_{K^{*0}K^{\mp}\pi^{\pm}}=3.76$ and $g_{\phi K^{+}K^{-}}=5.55$.
In the limit of SU(3) symmetry, we take
$g_{K^{*0}K^{\mp}\pi^{\pm}}=\sqrt{2}g_{K^{*\pm}K^{\mp}\pi^{0}}=\sqrt{6}g_{K^{*\pm}K^{\mp}\eta}$
and $\sqrt{2}g_{\omega(\rho^{0})
K^{\pm}K^{\mp}}=g_{\rho^{\pm}K^{\mp}K^{0}}=g_{\phi
K^{\pm}K^{\mp}}$. $g_{\rho^{\pm}K^{\mp}K^{*0}}=\sqrt{2}g_{\omega
K^{\pm}K^{*\mp}}=g_{\phi K^{*\pm}K^{\mp}}=6.48$ GeV$^{-1}$
\cite{zhao}. With $\Gamma(a_{1}\to K\bar{K}^{*}+c.c.)/\Gamma
({a_1})=(3.3\pm0.5\pm0.1) \% $ \cite{a1-exp} and $\Gamma
({a_1})=250\sim600$ MeV, we take the central value to roughly
estimate the coupling constant and get
$g_{_{a_{1}K^{\pm}\bar{K}^{*\mp}}}=0.63$ GeV \footnote{The decay
width of $a_{1}(1260)\to KK^{*}$ is written as $
\Gamma=\int^{\infty}_{(m_{K}+m_{K^*})^2}{\rm{d}}s
f(s,m_{a_{1}},\Gamma_{a_{1}})\frac{|\bf{k}|\;g_{_{a_{1}K^{\pm}\bar{K}^{*\mp}}}^{2}}{24\pi
s}\times\Big[2+\frac{{m_{K^{*}}^{2}}+|\bf{k}|^2
}{{m_{K^{*}}^{2}}}\Big]$, where the Breit-Wigner distribution
function $f(s,m_{a_1},\Gamma_{a_1})$ and the decay momentum
$|\bf{k}|$ are $ f(s,m_{a_1},\Gamma_{a_1})=\frac{1}{\pi}
\frac{m_{a_1}\Gamma_{a_1}}{(s-m_{a_1}^{2})^{2}+m_{a_1}^{2}\Gamma_{a_1}^{2}}$,
$|\bf{k}|=\frac{\sqrt{[s-(m_{K}+m_{K^{*}})^{2}][s-(m_{K}-m_{K^{*}})^{2}]}}{2\sqrt{s}}$.
}.

\begin{widetext}\begin{center}
\begin{figure}[htb]
\begin{center}
\begin{tabular}{cccccccc}
\scalebox{0.5}{\includegraphics{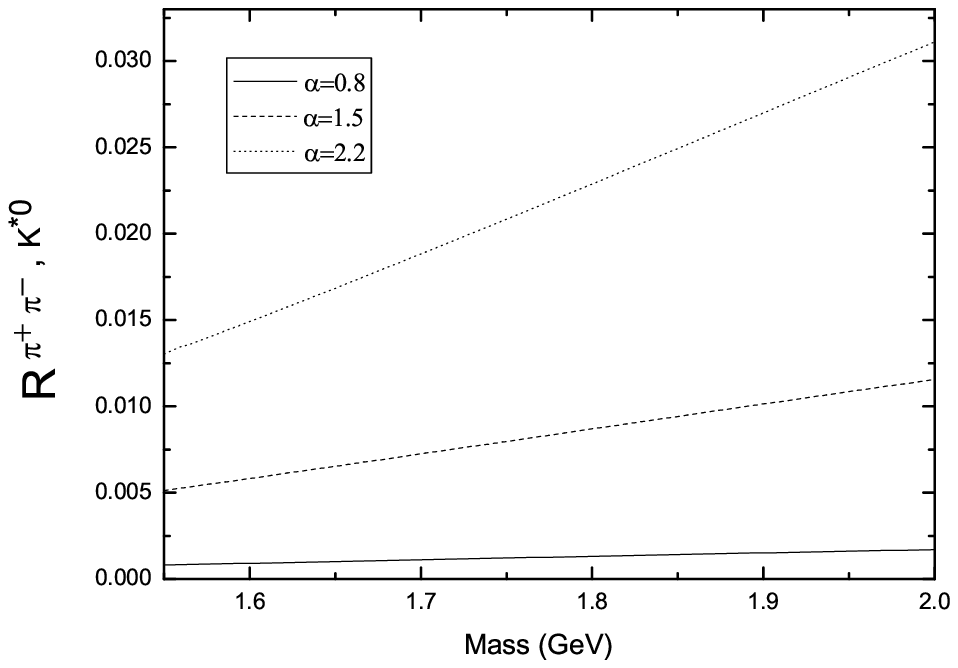}}&\scalebox{0.5}{\includegraphics{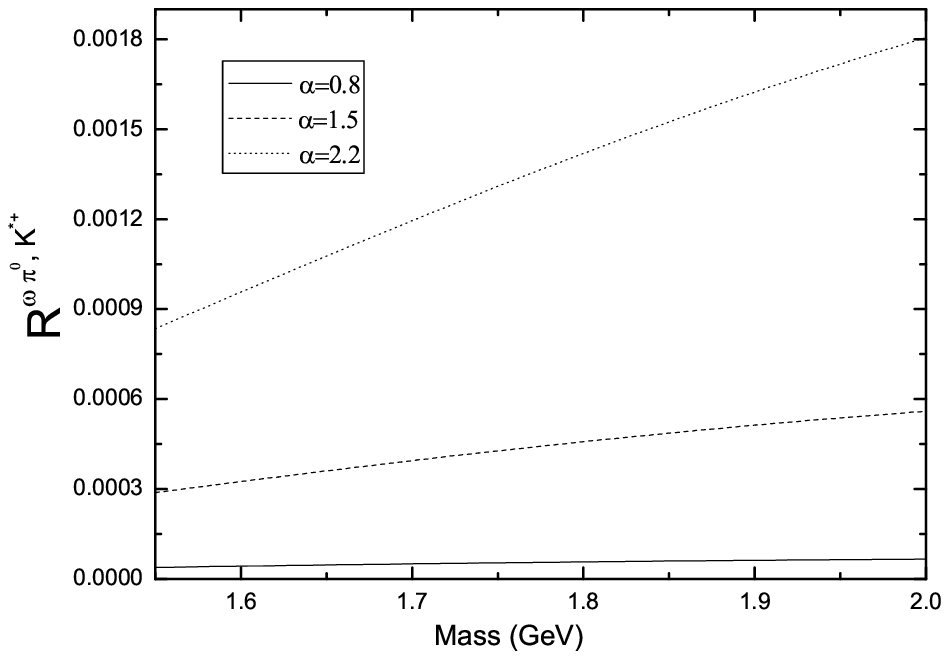}}&
\scalebox{0.5}{\includegraphics{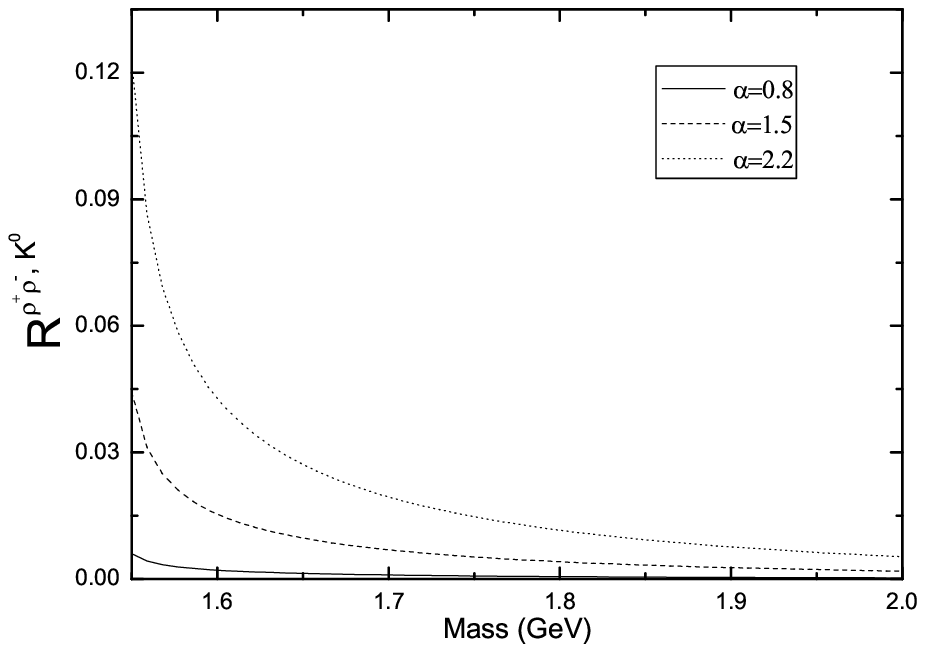}}\\
(a)&(b)&(c)\\
\scalebox{0.5}{\includegraphics{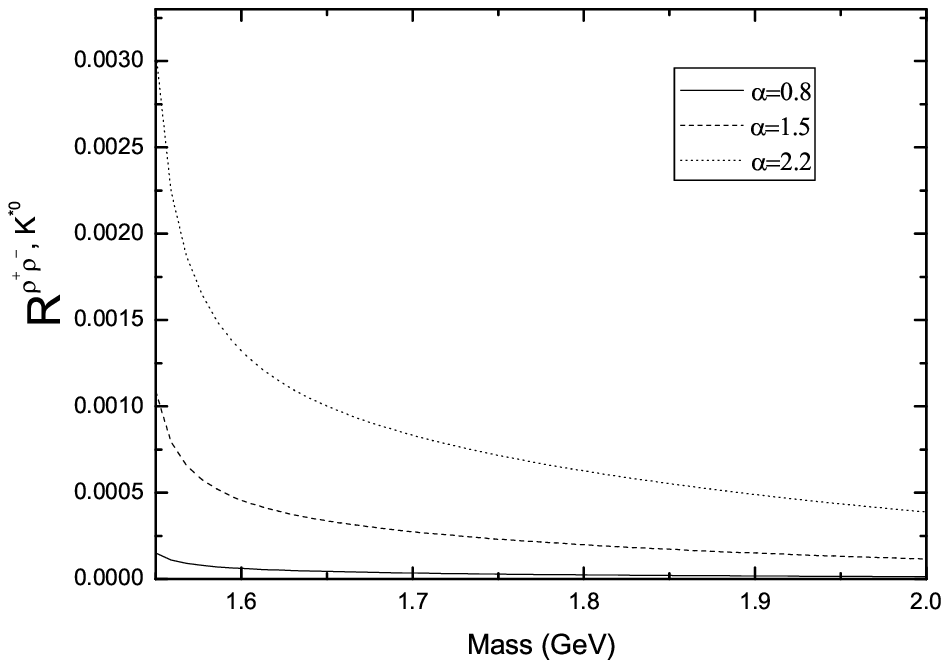}}&\scalebox{0.5}{\includegraphics{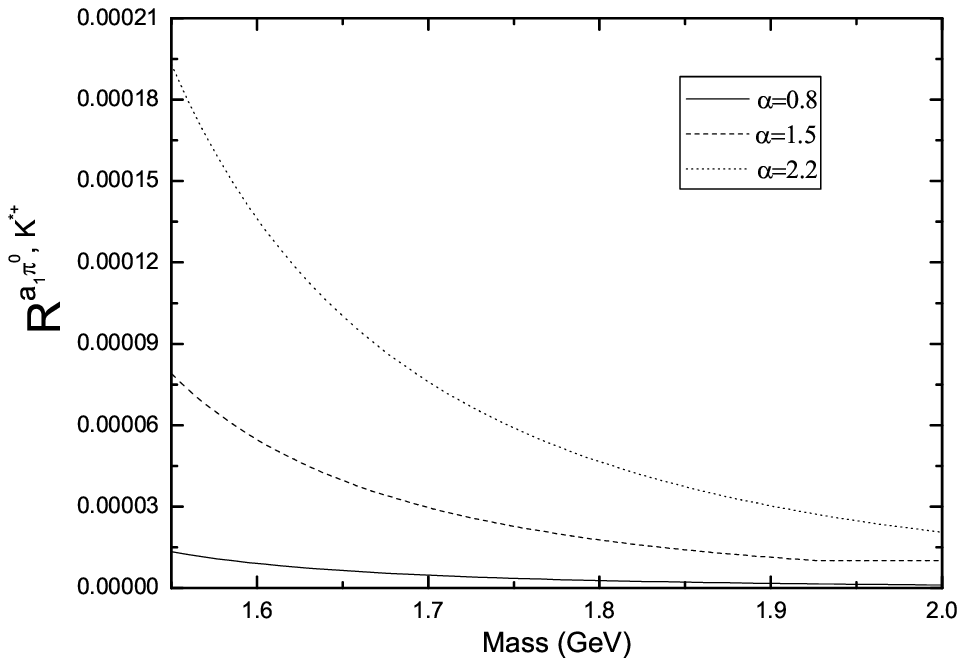}}&\\(d)&(e)&
\end{tabular}
\end{center}
\caption{The dependence of $R^{AB,C}$ corresponding to the
different intermediate states on the mass range of
$\rho(1450,1700)$ with monopole form factor. \label{fig2}}
\end{figure}\end{center}
\end{widetext}

\begin{figure}[htb]
\begin{center}
\scalebox{0.9}{\includegraphics{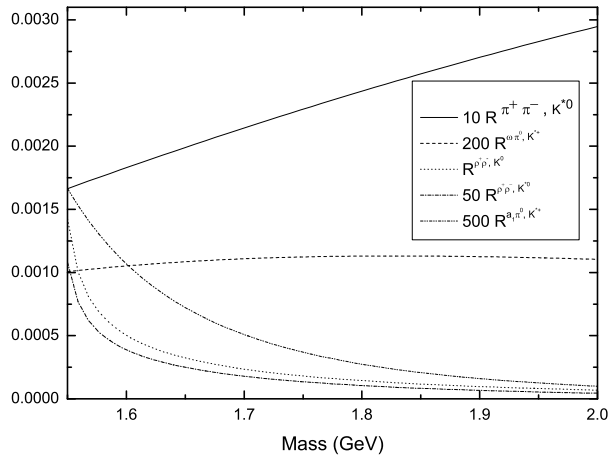}}
\end{center}
\caption{{ The dependence of $R^{AB,C}$ corresponding to the
different intermediate states on the mass range of $\rho(1450,1700)$
with the dipole form factor ($\alpha=1.5$).}\label{fig}}
\end{figure}

The dependence of $R^{AB,C}$ on the mass of $\rho(1450,1700)$ with
different intermediate states and $\alpha=0.8, 1.5, 2.2$ is shown
in Fig. \ref{fig2}, where the mass range of $\rho(1450,1700)$ is
$1.5\sim 1.8$ GeV. The re-scattering processes of $\pi^{+}\pi^{-}$
through exchanging a $K^*$ meson and $\rho^{+}\rho^{-}$ through
exchanging a $K$ meson contribute dominantly to
$\rho(1450,1700)\to K^{+}K^{-}$. $R^{AB,C}$ with intermediate
states $\rho^{0}\eta$ and $K^{*}K$ is of the same order as that
with the $\omega\pi^{0}$ intermediate state in the limit of SU(3)
symmetry, which is not shown in Fig. \ref{fig2}. For comparison,
we illustrate the variation of $R^{A,BC}$ with the mass of
$\rho(1450,1700)$ using the dipole form factor
$\mathcal{F}(m_{i},q^2)=[{(\Lambda^{2}-m_{i}^2
)}/{(\Lambda^{2}-q^{2})}]^2$ in Fig. \ref{fig}.

In the following we focus on the interference of the dominant
amplitudes corresponding to Fig. \ref{KK} (a) and (c), which reads
\begin{eqnarray}
&&\mathcal{M}\big[\rho(1450,1700)\to
K^{+}K^{-}\big]\nonumber\\&=&\mathcal{M}\big[\rho(1450,1700)\to
\pi^{+}\pi^{-}\to
K^{+}K^{-}\big]\nonumber\\&&+e^{i\phi}\mathcal{M}\big[\rho(1450,1700)\to
\rho^{+}\rho^{-}\to K^{+}K^{-}\big],\label{ampli}
\end{eqnarray}
where $\phi$ denotes the phase between Fig. \ref{KK} (a) and (c).
The dependence of the decay width from the above interference
amplitude on the parent mass and different $\phi$ is presented in
Fig. \ref{fig3} and Fig. \ref{fig4}.

Our numerical results indicate: (1) no enhancement structure
exists in the case of no interference as shown in Fig. \ref{fig2};
(2) the interference between Fig. \ref{KK} (a) and (c) leads to an
enhancement in a typical range of phase $\phi$ $120^{\circ}\sim
180^{\circ}$. Especially from Fig. \ref{fig4}, we notice that the
the enhancement with different phases occurs around $1540$ MeV,
very close to $X(1576)$. The enhancement depicted in Fig.
\ref{fig4} is similar to the cusp effect discussed in Ref.
\cite{bugg} to some extent. In fact such an enhancement occurs
with the opening of the $\rho\rho$ channel. Although the numerical
results depend on the particular parametrization of the form
factor, the qualitative features and conclusion remain essentially
the same.

The $K^+K^-$ spectrum from the above interference mechanism mimics
the observed broad spectrum from BES's measurement. However, basing
on the estimate from the calculation with monoploe form factor, the
decay width of $\rho(1450,1700)\to K^+ K^-$ from FSI effect is about
0.2 MeV only. Taking the width of $\rho(1450,1700)$ as 300 MeV, the
branching ratio of $\rho(1450,1700)\to A+B\to K^{+}K^{-}$ is about
$10^{-4}$. If the order of magnitude of $B[J/\psi\to
\pi+\rho(1450,1700)]$ is roughly $10^{-3}$ \cite{PDG}, the
$B[J/\psi\to \pi^{0}+\rho(1450,1700)]\cdot B[\rho(1450,1700)\to
AB\to K^{+} {K}^{-}]$ is about $10^{-7}$, which is far less than
experimental value $B[J/\psi\to \pi+X(1576)]\cdot B[X(1576)\to
K^{+}{K}^{-}]=8.5\times 10^{-4}$. The naive interpretation of the
extremely broad structure X(1576) arising from the final state
interaction effect seems not very favorable. Clearly more
experimental information on the exotic structure will be helpful.

\section*{Acknowledgments}
X.L. thanks Prof. X.Q. Li for useful discussion. This project was
supported by the National Natural Science Foundation of China
under Grants 10421503 and 10625521, Key Grant Project of Chinese
Ministry of Education (No. 305001) and the China Postdoctoral
Science foundation (No. 20060400376).

\begin{widetext}
\begin{center}
\begin{figure}[htb]
\scalebox{2.0}{\includegraphics{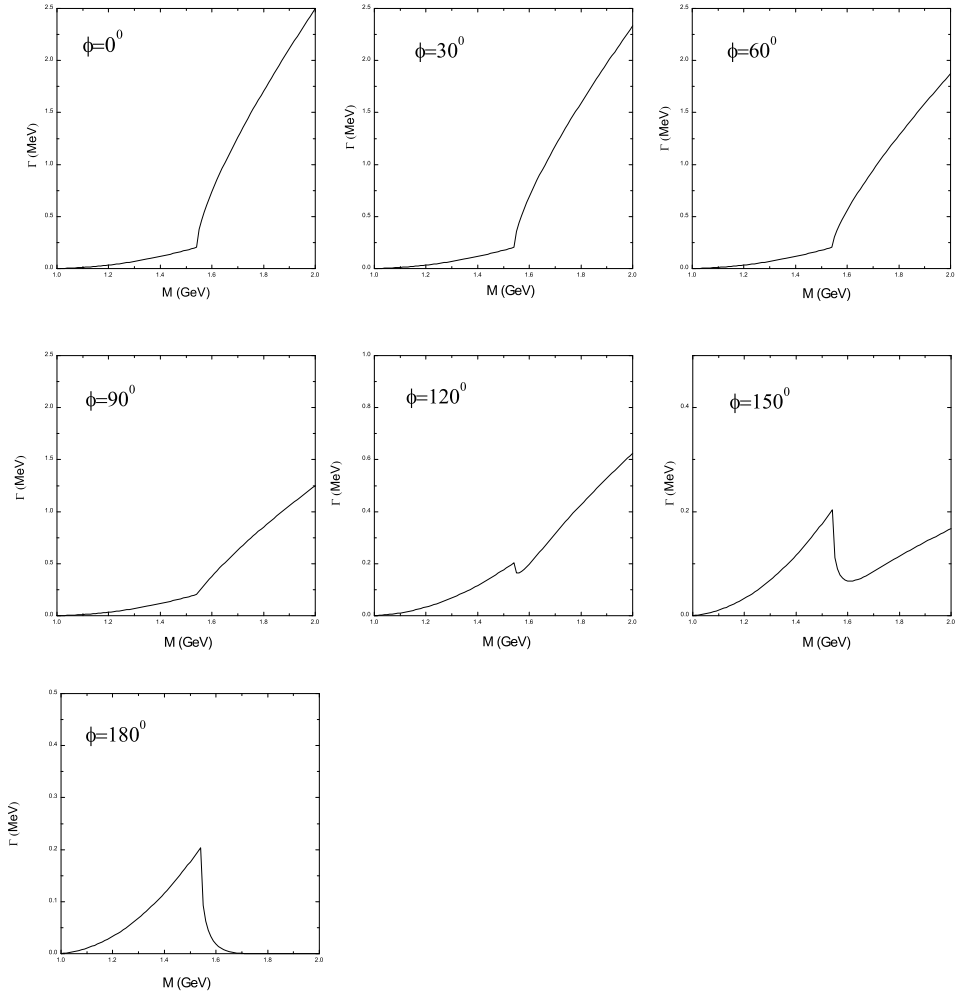}} \caption{The
dependence of the decay width from the interference amplitude
(\ref{ampli}) on the mass and $\phi$ using monopole form factor
with typical value $\alpha=1.5$ . \label{fig3}}
\end{figure}
\end{center}
\end{widetext}

\begin{widetext}
\begin{center}
\begin{figure}[htb]
\begin{tabular}{cc}
\scalebox{0.8}{\includegraphics{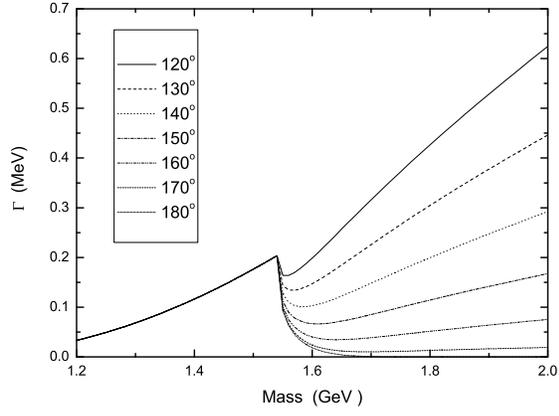}}&\scalebox{0.8}{\includegraphics{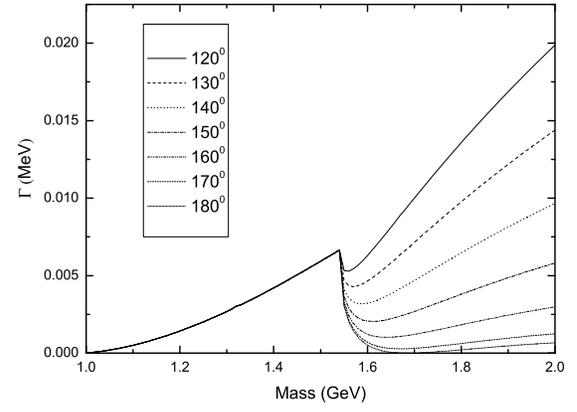}}\\
(a) &(b)
\end{tabular}
\caption{ (a) and (b) show the evolution of the decay width with
the mass and $\phi$ considering the monopole and dipole form
factors respectively with $\alpha=1.5$. \label{fig4}}
\end{figure}
\end{center}
\end{widetext}

\end{document}